# Superconductivity and Cobalt Oxidation State in Metastable $Na_xCoO_{2-\delta}\cdot yH_2O$ ($x \approx {}^1/_3$; $y \approx 4x$)

**P. W. Barnes, M. Avdeev*, J. D. Jorgensen, D. G. Hinks, H. Claus, and S. Short**

*Materials Science Division, Argonne National Laboratory, Argonne, IL 60439. E-mail: jjorgensen@anl.gov. * - Current Address: Bragg Institute, Building 58, Australian Nuclear Science and Technology Organisation, PMB 1, Menai NSW 2234, Australia*

**Abstract** We report the synthesis and superconducting properties of a metastable form of the known superconductor $Na_xCoO_2\cdot yH_2O$ ($x \approx {}^1/_3$; $y \approx 4x$). We obtained this metastable cobaltate superconductor due to the unique way it was synthesized. Instead of using the conventional bromine-acetonitrile mixture for the $Na^+$-deintercalation reaction, we use an aqueous bromine solution. Using this method, we oxidize the sample to a point that the sodium cobaltate becomes unstable, leading to formation of other products if not controlled. This compound has the same structure as the reported superconductor, yet it exhibits a systematic variation of the superconducting transition temperature ($T_c$) as a function of time. Immediately after synthesis, this compound is not a superconductor, even though it contains appropriate amounts of $Na^+$ and $H_2O$. The samples become superconducting with low $T_c$ values after ca. 90 h. $T_c$ continually increases until it reaches a maximum value (4.5 K) after about 260 h. Then $T_c$ drops drastically, becoming non-superconducting approximately 100 h later. Corresponding time-dependent neutron powder diffraction data shows that the changes in superconductivity exhibited by the metastable cobaltate correspond to slow formation of oxygen vacancies in the $CoO_2$ layers. In effect, the formation of these defects continually reduces the cobalt oxidation state causing the sample to evolve through its superconducting life cycle. Thus, the dome-shaped superconducting phase diagram is mapped as a function of cobalt oxidation state using a single sample. The width of this dome based on the formal oxidation state of cobalt is very narrow – approximately 0.1 valence units wide. Interestingly, the maximum $T_c$ in $Na_xCoO_2\cdot yH_2O$ occurs when the cobalt oxidation state is near +3.5. Thus, we speculate that the maximum $T_c$ occurs near the charge ordered insulating state that correlates with the average cobalt oxidation state of +3.5.

## Introduction

$Na_xCoO_2$ ($0.70 < x < 0.85$) and its $Na^+$-deintercalated analogues have been a major research interest in materials science, solid-state physics and chemistry over the past 20 years. Initially, this class of compounds was examined because of their technologically important properties, such as fast-ion conduction [1] for battery materials and large thermopower [2] for thermoelectric coolers. Recently, Takada *et al.* [3] reported novel superconductivity in oxidized, hydrated $Na_xCoO_2 \cdot yH_2O$ ($x \approx {}^1/_3$; $y \approx 4x$). This compound was synthesized from $Na_{0.7}CoO_2$ by oxidative deintercalation of $Na^+$ using a $CH_3CN/Br_2$ solution. Subsequent exposure to water formed the superconducting hydrated sodium cobaltate. $Na_{0.33}CoO_2 \cdot 1.3H_2O$ (space group $P6_3/mmc$; $a \approx 2.82$ Å; $c \approx 19.8$ Å) contains two-dimensional $CoO_2$ hexagonal planes formed from face-sharing $CoO_6$ octahedra. The sodium cations are coordinated by four $H_2O$ molecules forming distorted tetrahedra with the water molecules oriented in such a manner to form an extensive hydrogen bonding network throughout the lattice. The reader is referred to our previous work [4] that describes the crystal structure of $Na_xCoO_2 \cdot yD_2O$ ($x \approx {}^1/_3$; $y \approx 4x$) in detail.

Currently, the oxidization state of cobalt that supports superconductivity is not well defined. Schaak and co-workers [5] reported a superconductivity phase diagram based on varying the sodium content by changing the $Br_2$ concentration in their reaction mixture. All sodium cobaltate samples studied with sodium content ($x$) ranging from 0.26 to 0.45 were superconducting with transition temperatures ($T_c$) between 2 and 4.3 K. They showed that $T_c$ collected over the aforementioned sodium content range exhibited dome-shaped behavior. Based on the assumption that sodium content alone controlled the cobalt oxidization state, they suggested that the optimum $T_c$ occurred when $Co^{n+} = +3.7$. Milne *et al.* [6] were not able to reproduce these results. Using the same synthetic protocol, they found high $T_c$'s (> 4 K) over a broad range of sodium contents with only two samples showing low $T_c$ values. Using oxidation-reduction titrations to determine the oxidization state of their samples, they found that the maximum $T_c$ occurred near a cobalt oxidization state equal to +3.3. Additionally, the cobalt oxidation state in $Na_xCoO_2 \cdot yH_2O$ ($x \approx {}^1/_3$; $y \approx 4x$) and the parent material measured by titration performed in other publications have reported similar values well below what was expected based on sodium content alone. Such reports suggest that the Co oxidization state is set by defects other than the $Na^+$ concentration alone.

Two possible defects – proton incorporation into the interlayer spacing of sodium cobaltate hydrate and oxygen vacancy formation in the $CoO_2$ layers – have been suggested as possible causes for the apparent reduction in the cobalt oxidation state. Takada *et al.* [7] reported that $H_3O^+$ was incorporated between the $CoO_2$ layers of $Na_{0.33}CoO_2 \cdot 1.3H_2O$ without any additional loss of $Na^+$, thus lowering the Co oxidation state. However, Karppinen *et al.*

[8] suggested that the reduced oxidation state they determined for their samples by redox titration was due to the formation of oxygen vacancies in the cobalt oxide layers. Oxygen vacancy formation has also been suggested by Bañobre-López and co-workers [9] as the reason for the reduction in the measured cobalt valence.

In this paper we describe the synthesis and superconducting properties of a metastable $Na_xCoO_2{\cdot}yH_2O$. Samples made for this study were synthesized by oxidative deintercalation of the parent sodium cobaltate using an aqueous $Br_2$ solution instead of the more conventional acetonitrile solution. Other investigators have carried out this reaction in aqueous media using other oxidants. Park and co-workers [10] used a pH = 10.5 $Na_2S_2O_8/NH_4OH$ solution and Liu *et al.* [11] completed the reaction using $KMnO_4$. Although the $CH_3CN/Br_2$-based synthesis described above appears to be straightforward, the resultant products can be quite different even though they are reported to have the same sodium content. In addition, comparison of the previously published literature shows a wide variability in both the time needed to deintercalate $Na^+$ from $Na_xCoO_2$ ($x \approx 0.7$) and the time needed to hydrate the sodium-deficient sample to produce a sample with an optimum $T_c$ and bulk superconductivity (for examples, see refs. 5-9, 12). Using the synthetic method described in this paper, samples can be oxidized to a point that the material becomes prone to reduction by $H_2O$. We show that the final product of this reduction is CoOOH. This reduction to CoOOH is slow and the metastable product can be removed from the $H_2O/Br_2$ solution. After removal from the oxidizing solution, this metastable $Na_{0.33}CoO_2{\cdot}1.3H_2O$ is slowly reduced over a period of several days by formation of oxygen vacancies. The generation of oxygen vacancies leads to changes in the $T_c$ of the material. Initially, $T_c$ increases from values less than 2.0 K to 4.5 K over a 260 h period, then rapidly decreases below 1.4 K less than 100 h later. Therefore, we show that a single $Na_{0.33}CoO_2{\cdot}1.3H_2O$ sample can be used to trace the cobalt oxidation state dependence of the superconducting transition temperature. Based on our determination of oxygen vacancy concentration from neutron powder diffraction, we find that $T_c$ slowly increases from a cobalt oxidization state near +3.55(4) to a maximum near +3.48(4) before vanishing. Interestingly, the maximum $T_c$ is measured near $Co^{n+}$ = +3.5 where charge ordering is found to occur in the unhydrated base material.

## Experimental

Polycrystalline samples of the parent $Na_xCoO_2$ material ($x$ = 0.80-0.85) were synthesized by conventional solid-state methods. Appropriate amounts of $Na_2CO_3$ (Puratronic Alfa-Aesar, 99.997%; dried prior to use) and $Co_3O_4$ (Alfa-Aesar, 99.7% purity) were mixed intimately for 2 h in the presence of *n*-amyl alcohol using an agate ball mill. The samples were heated at 825–850°C for a total of 30 h in the presence of flowing $O_2$. The powders were reground and examined for sample purity by X-ray powder diffraction after each 10 h heating cycle in order to obtain high quality materials. Sodium loss due to volatility at elevated temperatures was small and was found by analysis to be about 2% for one of the samples. Table 1 lists the nominal sodium content for all the samples used in this investigation.

Oxidative deintercalation of $Na^+$ to form $Na_{0.33}CoO_2 \cdot 1.3H_2O$ was done using a $Br_2/H_2O$ solution, as reported in our previous paper [4]. We generically refer to any water-based chemistry as using $H_2O$ although any sodium cobaltate samples made specifically for neutron diffraction were synthesized using $D_2O$ (Alfa Aesar, 99.8% Isotopic). The solution used for the reaction contained ratios of $Br_2$-to-anhydrous sodium cobaltate and $H_2O$-to-$Na_xCoO_2$ as shown in Table 1. The bromine, water, and $Na_xCoO_2$ were continually mixed for 4 h before separating the resultant powder from the solution by filtration. The sample was washed three times with $H_2O$ to remove any residual bromine trapped in the wet powder. Excess water was removed from the resultant powder by storing it in a dessicator containing a saturated $KBr/H_2O$ (or $D_2O$) solution (58% relative humidity) for four days. Formation of $Na_xCoO_2 \cdot yH_2O$ was confirmed by X-ray powder diffraction.

The kinetics of $Na^+$ deintercalation from the parent sodium cobaltate were measured by removing small aliquots of the powder from a stirred reaction mixture for sample #1. The Na:Co mole ratios for all aliquots measured from sample #1 were determined by inductively coupled plasma optical emission spectroscopy (ICP-OES). Each sample was weighed and dissolved in aqua regia before analysis. Measurements were done by comparing the Na/Co signal intensity ratios from the samples with corresponding ratios from standards prepared with known Na:Co mole ratios. The measurements based on the Na/Co standard solutions allow finer distinction to be made between samples that have close-together element ratios.

Superconducting transition temperatures were determined for sample #4 from magnetization (M) vs. temperature (T) curves measured on two different magnetometers. Initially, small samples were removed periodically to measure $T_c$ on a Quantum Design PPMS 6000 magnetometer. After 4 days, a portion of the sample was hermetically-sealed in a small demagnetized stainless steel capsule in order to hold the water content of the sodium cobaltate hydrate constant. The free volume inside the sealed stainless steel container was not large enough to allow significant changes in the water content of $Na_{0.33}CoO_2 \cdot 1.3H_2O$. Field cooled magnetization measurements (H = 1 G) were made on sealed sample #4 using a non-commercial SQUID magnetometer described elsewhere [13]. Temperatures below 4.2 K were reached by pumping on liquid helium condensed into the sample chamber.

The remainder of sample #4 and all of sample #5 were used for neutron powder diffraction structural studies as a function of time. A series of time-dependent neutron diffraction patterns were collected on these samples at the Special Environment Powder Diffractometer at the Intense Pulsed Neutron Source [14] to determine what possible structural changes occur with time as the $T_c$ of the sample changes. No diffraction data were taken for the first four days while sample #4 was stored in a dessicator containing a saturated $Ca(NO_3)_2/D_2O$ solution (50% relative humidity). This was done in order to ensure that no free $D_2O$ was in the sample. Sample #5 was dried in a $N_2$-filled glove bag and subsequently rehydrated over a saturated $KNO_3$ solution (92% relative humidity) for 24 h. Each sample was sealed in a vanadium can to prevent changes in the water content of the sodium cobaltate deuterate and diffraction data was collected periodically. Structural analysis based on the model we previously reported [4] was completed on high resolution time-of-flight neutron diffraction data (detector bank #1; $2\Theta = 144°$) using the Rietveld method as implemented in the EXPGUI/GSAS software package [15, 16].

## Results and Discussion

### 1. *Formation of metastable* $Na_xCoO_2•yH_2O$

The kinetics of the deintercalation reaction measured by ICP-OES on sample #1 is shown as a function of time in Fig. 1. Formation of $Na_{0.33}CoO_2·1.3H_2O$ is essentially complete after 1-2 h, but as shown in the Fig. 1 inset, there is a very slow decrease in sodium content with additional time. The exchange of $Na^+$ by $H_3O^+$ as reported in ref. 7 is a plausible explanation. However, as we will show later, the slow sodium loss is due to the reduction of $Na_{0.33}CoO_2·1.3H_2O$ by $H_2O$.

After long reaction times and/or large initial $Br_2$ concentrations, a secondary phase is formed as shown by the partial X-ray diffraction patterns in Fig. 2a. A broad peak develops at ca. 19° $2\theta$ 24 h after the beginning of the deintercalation reaction, near the 004 reflection of $Na_{0.33}CoO_2·1.3H_2O$. This reflection increases in intensity as the reaction proceeds over longer time periods. Yet, even after 72 h, we were unable to determine what this poorly formed impurity phase was because its phase fraction was not significant enough to observe additional reflections related to this phase in the diffraction data. In order to determine what this unknown phase was, we allowed sample #2c to react with the bromine water solution for two weeks. Fig. 3a shows the result of this reaction after the product was dried by flowing nitrogen. Additional broad reflections were observed at ca. 41.7° and 55.4° $2\theta$ and the intensity of the 100 peak had increased. This diffraction pattern was indexed and the two phases present were identified as $Na_{0.33}CoO_2·0.6H_2O$ and CoOOH. $Na_{0.33}CoO_2·0.6H_2O$ is another known sodium cobaltate hydrate which has a structure related to the superconducting sodium cobaltate [7]. Instead of forming $Na(H_2O)_4^+$ tetrahedra between the $CoO_2$ layers, the water molecules reside in the same plane as the sodium cations.

CoOOH, also known as heterogenite by mineralogists, exists as two different polymorphs – a hexagonal (2H) form and a rhombohedral (3R) form [17]. Our X-ray diffraction data is consistent with 2H-CoOOH. The 2H structure has the same base structure as $Na_{0.33}CoO_2 \cdot 1.3H_2O$ with only $H^+$ separating the $CoO_2$ planes. The $CoO_2$ layers in the 3R-form of the cobalt oxyhydroxide have a different periodicity than seen in the 2H form. Chemical synthesis generally leads to the 3R-polymorph; however, since the structure of the 2H form is closely related to that of $Na_{0.33}CoO_2 \cdot 1.3H_2O$, it is probable that the room temperature soft chemical synthesis would lead to 2H-CoOOH.

Additionally, the lattice parameters calculated from the three strongest diffraction peaks shown in Fig. 3a are $a \approx 2.84$ Å and $c \approx 9.12$ Å. Deliens and Goethals [17] reported $a = 2.855$ Å and $c = 8.805$ Å for the mineral heterogenite. We find a much longer $c$ lattice parameter than the previously reported values. The difference between the unit cell constants reported here and those reported by Deliens and Goethals must be attributed to the different forms of the CoOOH samples examined. The previous work focused on a well crystallized geological sample. In this work, the poorly crystallized cobalt (III) oxyhydroxide sample was made through a solution-based route at room temperature and its possible that other products from the decomposition of $Na_{0.33}CoO_2 \cdot 1.3H_2O$ may also be present. However, we are confident in the assignment of 2H-CoOOH as the end point of this reaction due to reasonable agreement between the previously reported $a$ lattice parameter and the one reported here, which reflects the cobalt oxidation state, and that moderate heating of this sample (325°C) gave a mixed phase material containing $Na_xCoO_2$ formed from dehydration of $Na_{0.33}CoO_2 \cdot 0.6H_2O$ and $Co_3O_4$ formed from the decomposition of CoOOH (see Fig. 3b) [18].

Next, we examined the amount of CoOOH formed during deintercalation as a function of increasing $Br_2$ content. As shown in Fig. 2b, when the $Br_2$ concentration was increased, the amount of CoOOH formed increased given that all other variables, such as reaction time, were fixed. For very large $Br_2$ concentrations, the kinetics of CoOOH formation is independent of bromine content as shown in Fig. 2c for sample #3.

From the data presented above, we can make the following generalizations about the chemistry which occurs during formation of metastable $Na_{0.33}CoO_2 \cdot 1.3H_2O$. First, as oxidization of $Na_xCoO_2$ proceeds, it eventually reaches a point where it is no longer chemically stable with respect to reduction by $H_2O$. It then reacts either with water intercalated in between the $CoO_2$ layers or that from oxidative medium as shown in equation **1**.

$$2\ Na_xCoO_2 \cdot yH_2O \longrightarrow 2\ CoOOH + 2x\ Na^+ + 2x\ OH^- + (2y-x-1)\ H_2O + (1-x)/2\ O_2 \qquad \textbf{(1)}$$

One would expect that the amount of $Br_2$ used would affect the overall reaction until an appropriate amount was added to keep the solution saturated during oxidization. However, during the reaction, $Br^-$ is formed and reacts with excess $Br_2$ as shown in equation **2** to increase the solubility of bromine in the oxidizing solution.

$$Br_{2\ (aq)} + Br^- \longrightarrow Br_3^- \qquad \textbf{(2)}$$

The total oxidizing power of the solution is the combination of the concentrations of the aqueous bromine and tribromide ion. Using the known weights of the parent $Na_xCoO_2$, $Br_2$, and $H_2O$ in the reaction mixture, we can calculate the final combined $[Br_2 + Br_3^-]$ concentration. We assumed that the solubility of liquid bromine (0.214 mol $L^{-1}$) in water was unaffected by the presence of $Br^-$ and $Br_3^-$ in the solution. In addition, it should be noted that the difference between $Br_2$ solubility in $H_2O$ and $D_2O$ are ignored because such data for $D_2O$ are not readily available in the literature. The moles of $Br_2$ required to saturate the solution ($n_s$), leading to maximum oxidization of parent sodium cobaltate is:

$$n_s = 1.28 n_o (x - 0.33) + 0.214V \qquad \textbf{(3)}$$

where $n_o$ is the moles of $Na_xCoO_2$ ($0.80 \leq x \leq 0.86$) used and $V$ is the volume of water added to the reaction mixture. The ratio of the moles of added bromine to $n_s$ is used to indicate the oxidization potential of the reaction mixture and is reported in Table 1. Samples made in solutions with oxidation potentials greater then unity should all have the same reaction kinetics. Fig. 2c shows this effect with three samples with large excess quantities of $Br_2$.

$T_c$ and structural investigations were done on $Na_xCoO_2{\cdot}yH_2O$ samples made using an aqueous $Br_2$ solution with an oxidation potential around 0.65. This would be equivalent to adding enough $Br_2$ to make a saturated solution and to reduce the sodium content to $x \approx 0.33$. Note that Fig. 2a shows the time evolution of the CoOOH formation for a sample with this exact reaction ratio. After 6 h, CoOOH is first observed, so stopping the reaction at 4 h gives metastable $Na_xCoO_2{\cdot}yH_2O$ with cobalt in its highest possible oxidization state. This is how samples #4 and #5 were made for superconductivity and structural studies.

### 2. *Superconductivity and structural studies of metastable $Na_xCoO_2{\bullet}yH_2O$*

Figs. 4 and 5 show the superconducting transitions and $T_c$ determined as a function of time for sample #4, respectively. Because of the broad superconducting transitions (typical of all reported measurements for this material), we used the onset $T_c$ defined as the temperature at which M ≈ -0.25. The time elapsed for each sample was measured from when filtration was complete in the synthetic protocol. The times reported here reflect aging of the sample at room temperature. Any time contribution in which the sample was held at low temperature for magnetization measurements was subtracted from the overall time after filtration. Initially, no measurable $T_c$ was observed. Superconductivity ($T_c$ = 2.44 K) was first measured after 88 h. $T_c$ continued to increase until it reached a maximum value near 4.5 K approximately 260 h after filtration was completed. After $T_c$ peaked, it rapidly decreased below the SQUID's measurable temperature limit (1.4 K) in less than 100 h.

Over the same period as M vs. T measurements were taken, neutron diffraction data was collected simultaneously in order to probe for any structural changes that occurred as the sample's superconductivity was changing. The lattice constants of the metastable compound continually changed with time – *a* slightly increased and *c* contracted significantly as shown in Fig. 6a. The initial rapid change in the lattice parameters slowed with time but over the time scale we investigated, they never fully stabilized. Fig. 6b shows the change in *c*/*a* and $T_c$ as a function of time for sample #5. Superconductivity is observed only over the small, rapidly changing region of *c*/*a*. Rietveld refinements of the neutron diffraction data revealed the crystallographic parameter that causes the continuous change in the lattice parameters. The change in *c*/*a* occurs in response to the formation of oxygen vacancies in the $CoO_2$ planes, as shown in Fig. 7. The vacancy concentrations determined here from refined oxygen site occupancies are admittedly small, but time dependence of their formation scales nicely with changes in *c*/*a*. The fact that these two structural parameters scale with one another over a long time period gives us confidence that the refined vacancy concentrations are meaningful. In addition, no CoOOH is observed in the diffraction data even after long storage times. The reduction mechanism is either much different between the dried powder and the sample in the oxidizing medium or oxygen vacancy formation is the first step in the reduction to make cobalt (III) oxyhydroxide.

Fig. 8 shows the oxygen site occupancy as a function of time for samples #4 and #5 identically oxidized to $Na_{0.33}CoO_2{\cdot}1.3H_2O$. The fundamental difference between samples #4 and #5 is that sample #4 that was never dehydrated whereas sample #5 was dried in a nitrogen-filled glove bag for approximately 90 h before rehydration. The time dependence of the vacancy content has the same functional form for both samples but the time scale for sample #5 must be shifted by approximately 50 h to bring both curves into registry. Thus, the oxygen vacancy formation seen in $Na_{0.33}CoO_2{\cdot}0.6H_2O$ must be significantly slower than that seen in $Na_{0.33}CoO_2{\cdot}1.3H_2O$.

The oxidization states of cobalt can be determined based on a few simple assumptions about the initial state of the sample and its change in composition as a function of time. First, the fact that the oxygen site occupancy values are greater than 1, as shown in Figs. 7 and 8, cannot be real and suggests that there are some correlations in the refinements related to this parameter. Therefore, the site occupancy values reported here are not accurate. The large systematic error seen in refinement parameters such as the oxygen site occupancy is likely due to the diffuse scattering observed in the neutron diffraction data. The diffuse scattering arises from an ordered supercell of the sodium cobaltate hydrate, as discussed in Ref. 4.

However, the trend in site occupancy changes, as shown in Fig. 8, should be viewed as accurate since refinement of all the neutron diffraction data were handled identically. Since no diffraction data were collected during the 100 h after filtration, the initial site vacancy value is not known and extrapolation of the curve is not feasible since the site occupancy is rapidly changing. An additional sample of $Na_{0.33}CoO_2{\cdot}1.3H_2O$ was synthesized and dehydrated rapidly in an effort to determine an initial oxygen site occupancy. A sample was oxidized, filtered and freeze-dried within 20 h. Since the sample was changing rapidly, neutron powder diffraction patterns could not be collected for lengthy time periods to give an appropriate signal-to-background ratio in order to reasonably refine the oxygen site occupancy. However, $c/a$ was accurately determined from the data in spite of the poor statistics. This sample showed a very large $c/a$ value of 7.054(1). For comparison, the $Na_{0.33}CoO_2{\cdot}1.3H_2O$ sample with the largest $c/a$ value currently reported in the literature is 7.008 [6] and all reported values of $c/a$ fall between 6.94 and 7.01. By extrapolation of the data in Fig. 7, the oxygen site occupancy for this sample was determined to be ca. 1.032(2).

The second assumption is that the initial oxygen site occupancy is 1.00 immediately after synthesis. The parent $Na_xCoO_2$ sample was synthesized in flowing $O_2$, which should prevent vacancy formation in the starting material [18]. We assume that during the rapid synthetic procedure, no oxygen vacancies are formed. Normalizing the site occupancy by 1.032(2) then allows the determination of vacancy content with time. Fig. 7 shows this calculated vacancy content ($\delta$) and how it reaches values slightly greater than 0.1 after long time periods. Values of $\delta$ reported in this work are in excellent agreement with those values reported in previously published reports [8, 9]. The cobalt oxidization state is then calculated from the known sodium content of 0.33 and the vacancy content determined from the normalized site occupancy using the curve shown in Fig. 8.

Fig. 9 shows that the calculated Co oxidization state exhibits an asymmetric dome-like behavior where the total width of the dome is approximately 0.1 valence units. Starting from a higher formal valence, the material is a bulk superconductor as $T_c$ increases from sub-2 K temperatures to the maximum of 4.5 K. Once the sample reached a maximum $T_c$, the measured susceptibility decreased rapidly as $T_c$ decreased on the other side of the dome. Note that the position of the dome is determined by how one estimates the oxygen site occupancy at zero time. The error in this parameter is large, leading to an uncertainty in the dome position as a function of cobalt oxidation state (±0.04 valence units). The shape of the dome is more robust since it is determined by the shape of the curve in Fig. 8. The phase diagram suggested here is similar to the one reported by Schaak *et al.* [5], having about the same dome-shaped width as a function of cobalt valence, but with a much more asymmetric shape. Milne *et al.* [6] proposed a slow increase in $T_c$ with decreasing oxidization state but show no data on the low valence side. It should be noted that they find high transition temperatures over a wide range of oxidization state. In fact the presence of dome shaped behavior is not clear in their work. Our maximum $T_c$ occurs at an oxidization state between both of these works.

Our results can be interpreted in two ways. First, the behavior could be like that of the layered cuprate superconductors where there is bulk superconductivity in both the under- and over-doped regimes. However, the loss of superconductivity in our sample appears to occur very sharply upon passing through the maximum $T_c$ rather than scaling parabolically with the defect concentration.

A second possible explanation is that these materials do not display a full superconducting dome like the layered cuprates. Instead, the superconducting transition temperature may increase with changing carrier concentration until the maximum $T_c$ is reached, at which point, some other instability destroys superconductivity. Such behavior is often seen in conventional superconductors like the Chevrel phases [19] and chemically substituted $BaBiO_3$.[20] For such a scenario, the weak indications of superconductivity in the low-oxidization state doped regime beyond the optimum Co oxidation state may simply be the result of chemical inhomogeneity in the sample.

Such behavior could be attributed to the fact that the optimum $T_c$ and its sharp decrease occurred near $Co^{n+} = +3.5$. This raises the possibility that superconductivity is destroyed by charge ordering. Charge ordering in the base $Na_xCoO_2$ material occurs at $x = 0.5$, giving rise to an insulating phase that has been extensively studied [21]. Recently, Chen and co-workers [22] observed in $Na_{0.41}CoO_2$ was reduced in $H_2O$ to achieve a charge ordering with a cobalt oxidation state equal to +3.5. This charge ordered state was reached by the incorporation of $H_3O^+$ ions, similar to the case presented here where oxygen vacancies provided the reduction. The neutron powder diffraction showed no major change when superconductivity was lost which is not surprising since observing charge ordering by diffraction is difficult.

## Conclusion

Highly oxidized, metastable $Na_{0.33}CoO_2 \cdot 1.33H_2O$ can be directly formed from an aqueous $Br_2$ solution through proper adjustment of bromine content and reaction time. In a saturated solution, sodium is rapidly removed from the parent $Na_xCoO_2$ ($0.80 < x < 0.86$) until $x \sim 0.33$ is reached. Continuation of the reaction does not reduce the $Na^+$ content further. Instead, the $Co^{4+}$-containing sodium cobaltate hydrate is slowly reduced by water to CoOOH. Removal of $Na_{0.33}CoO_2 \cdot 1.33H_2O$ from the $Br_2/H_2O$ solution by filtration stops this reduction; however, the material is still unstable due to the presence of $Co^{4+}$ in the $CoO_2$ planes.

Initially this material showed no $T_c$, but with time superconductivity appeared. Powder neutron diffraction showed the formation of oxygen vacancies over this time period that slowly reduced the formal valence of cobalt. We found a very asymmetric dome for $T_c$ determined as a function of oxidation state, with the first hints of superconductivity starting at $Co^{+3.6}$. $T_c$ slowly increased over 260 h to its maximum value (4.5 K) near $Co^{3.5+}$, then within 100 h later, no superconductivity was observed. We believe that the rapid loss of superconductivity near this cobalt oxidation state may be due to charge ordering. This charge localization, previously observed in the anhydrous sodium cobaltates, would destroy superconductivity.

These time-dependent experiments provide a novel way to map the superconducting phase diagram as a function of time using a single sample. Additionally, these results show that the possibility of time-dependent phenomena should be considered in any experiments on $Na_xCoO_2 \cdot yH_2O$. In particular, for a given composition, one cannot be certain that the maximum $T_c$ has been reached without waiting to see whether the properties change with time. Even more troublesome, samples that show no superconductivity may simply have gone through their "life cycle" before being studied.

**Acknowledgements** This work is supported by the U.S. Department of Energy, Basic Energy Sciences – Materials Sciences, under Contract No. W-31-109-ENG-38. The authors would like to thank D. Graczyk, A. Essling, and S. Naik from the Analytical Chemistry Laboratory, Chemical Engineering Division at Argonne National Laboratory for performing the ICP-OES analysis. The authors would also like to thank A. Cheetham at the University of California–Santa Barbara for helpful discussions during the completion of this work.

**Table 1** Reaction conditions used to make $Na_{0.33}CoO_2 \cdot 1.3H_2O$ materials for this investigation. The initial Na content ($x$) of the parent material ($Na_xCoO_2$), g $Br_2$ used/g $Na_xCoO_2$, g water added/g $Na_xCoO_2$ and moles ($n$) of added $Br_2$/moles of $Br_2$ needed to saturate the solution ($n_s$) are given for each oxidization reaction. Sample numbers marked with an asterisk indicate that the sample was made using $D_2O$ instead of $H_2O$.

| **Sample number** | $x$ | **g $Br_2$/g $Na_xCoO_2$** | **g water/g $Na_xCoO_2$** | $n$ **($Br_2$)/**$n_s$ |
|---|---|---|---|---|
| 1* | 0.80 | 1.07 | 22.5 | 0.65 |
| 2a | 0.85 | 0.51 | 19.0 | 0.32 |
| 2b | 0.85 | 1.07 | 19.8 | 0.65 |
| 2c | 0.85 | 2.71 | 20.7 | 1.61 |
| 3a* | 0.86 | 2.76 | 23.2 | 1.55 |
| 3b* | 0.86 | 5.55 | 24.5 | 3.04 |
| 3c* | 0.86 | 9.96 | 23.0 | 5.61 |
| 4* | 0.80 | 1.04 | 21.6 | 0.64 |
| 5* | 0.80 | 1.04 | 21.5 | 0.64 |

**Figure 1** The measured Na:Co ratio as a function of reaction time for sample #1 undergoing deintercalation. The inset shows the gradual decrease in Na:Co ratio after the reaction is essentially complete. The lines shown here are drawn to guide the eyes.

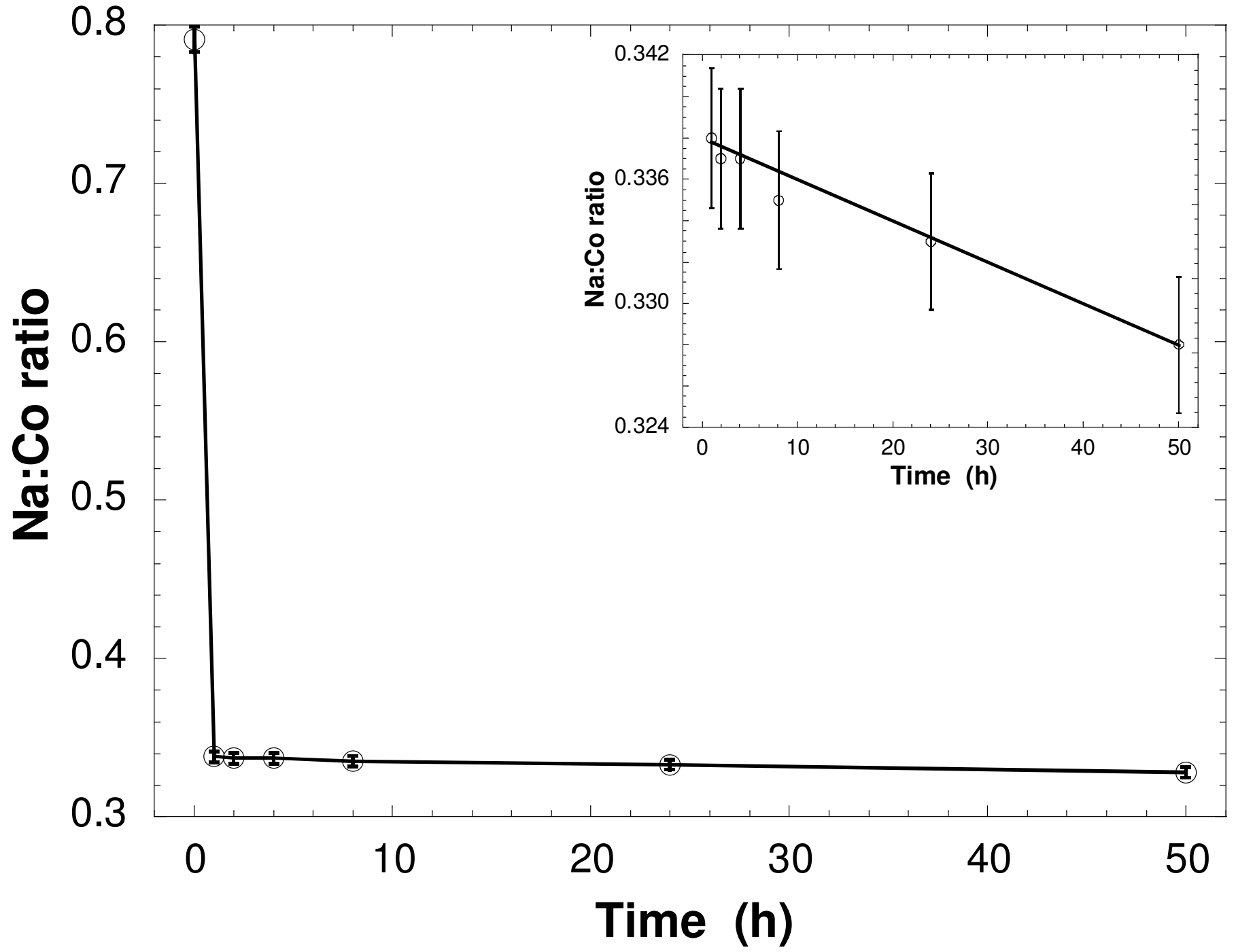

**Figure 2** X-ray diffraction data (a) collected as a function of reaction time for sample #2b; (b) collected as a function of $Br_2$ concentration for 3 different reaction mixtures ($n$ ($Br_2$)/$n_s$ given in parentheses) for sample #2; (c) showing the effect of added $Br_2$ above the point where liquid bromine is present for samples #3a, b and c with 1.55, 3.04 and 5.61 $n$ ($Br_2$)/$n_s$, respectively. In all diffraction patterns the $Na_{0.33}CoO_2{\cdot}1.3H_2O$ 004 peak is normalized to 1000 counts.

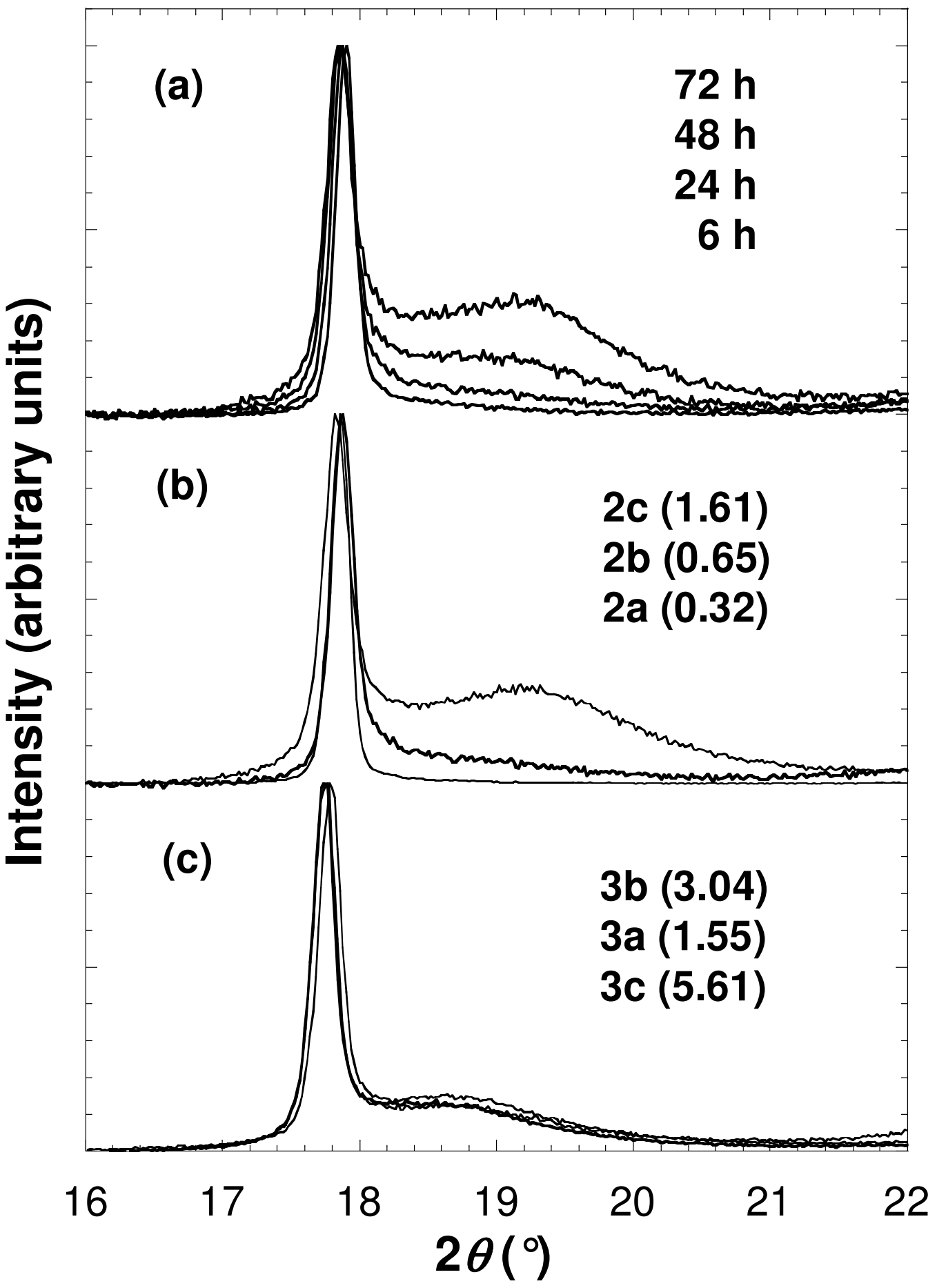

**Figure 3** X-ray diffraction patterns of sample #2c after (a) 2 weeks reaction time and (b) heated to about 325°C. Legend: Unmarked *hkl* values represent (a) $Na_{0.33}CoO_2 \cdot 0.6H_2O$ and (b) unhydrated $Na_xCoO_2$. Underlined *hkl* values denote (a) CoOOH and (b) $Co_3O_4$.

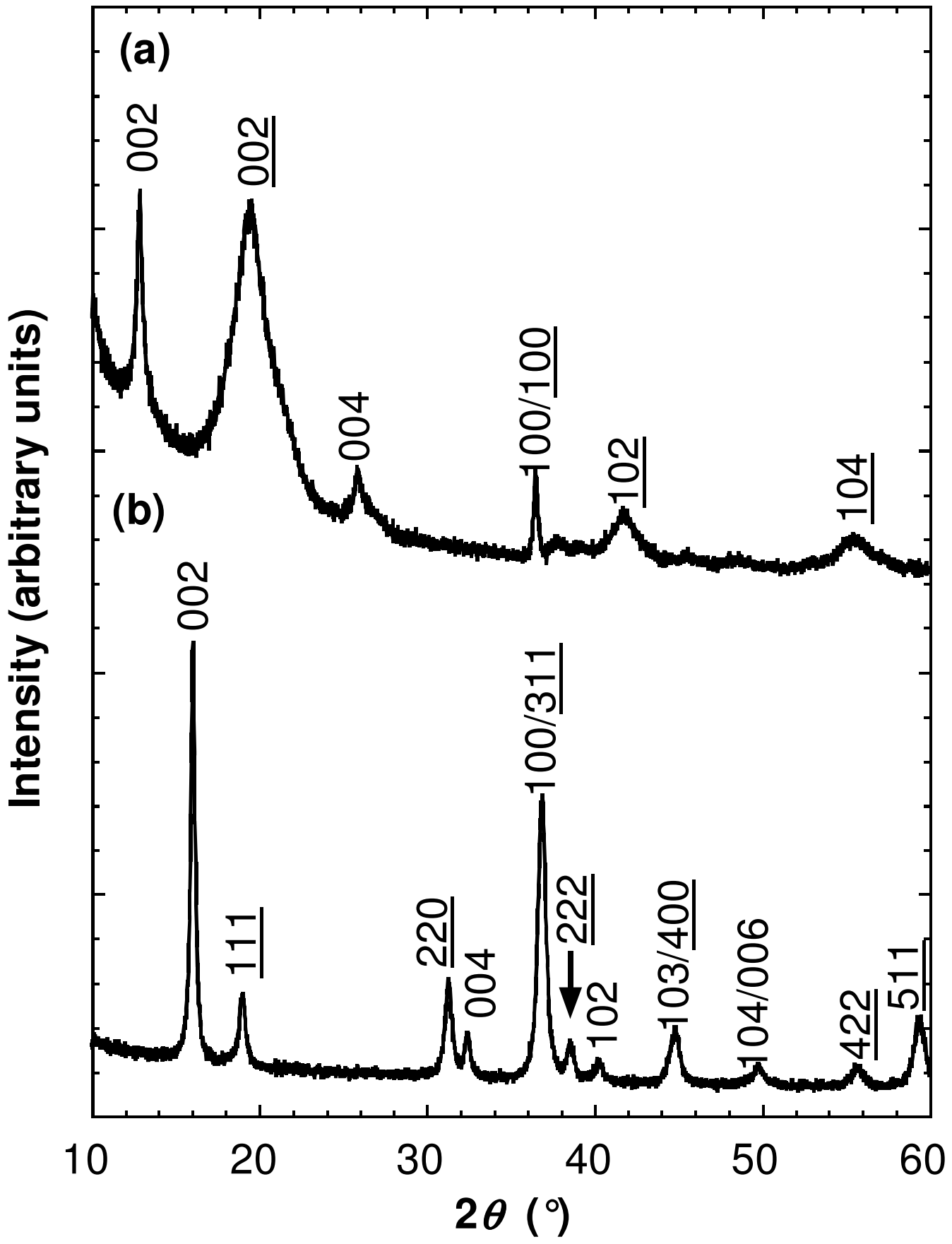

**Figure 4** M vs. T curves collected on a non-commercial SQUID magnetometer for sample #4. The upper panel shows curves as $T_c$ (onset) increases with time whereas the lower panel represents decreasing $T_c$ with increasing time. The same sample sealed in a non-magnetic stainless steel container was used for all measurements shown here.

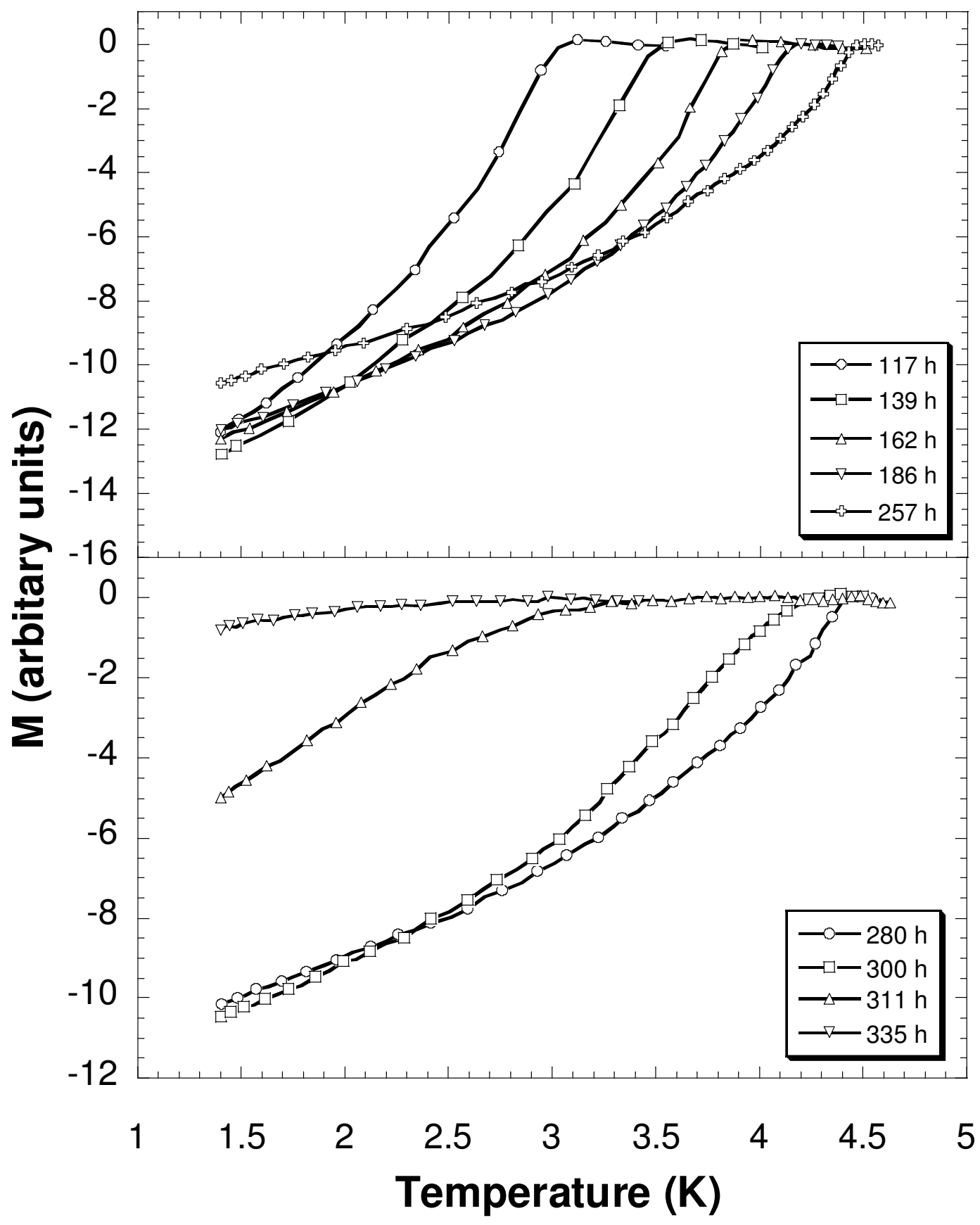

**Figure 5** $T_c$ (onset) vs. time for sample #4 kept at room temperature. Legend: circles – $T_c$ determined from SQUID data; squares – $T_c$ determined from PPMS data. The line shown here is drawn to guide the eyes.

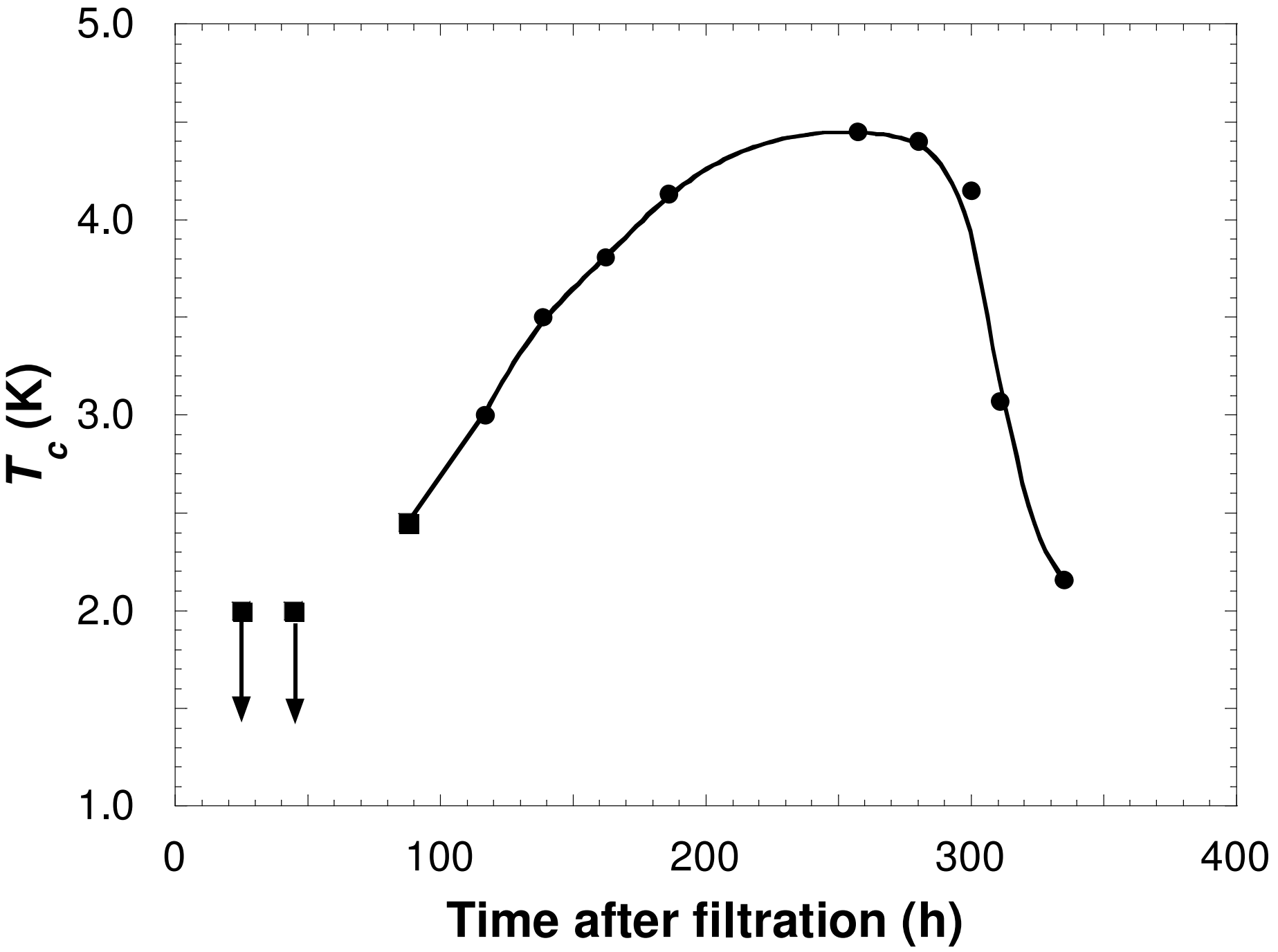

**Figure 6** Lattice parameters and $T_c$'s for sample #5 vs. time after synthesis. The top panel shows the *a* (circles) and *c* (squares) unit cell constants for $Na_{0.33}CoO_2{\cdot}1.3H_2O$. The bottom panel shows changes in *c*/*a* and $T_c$ for $Na_{0.33}CoO_2{\cdot}1.3H_2O$ as a function of time. All lines shown here are drawn to guide the eyes.

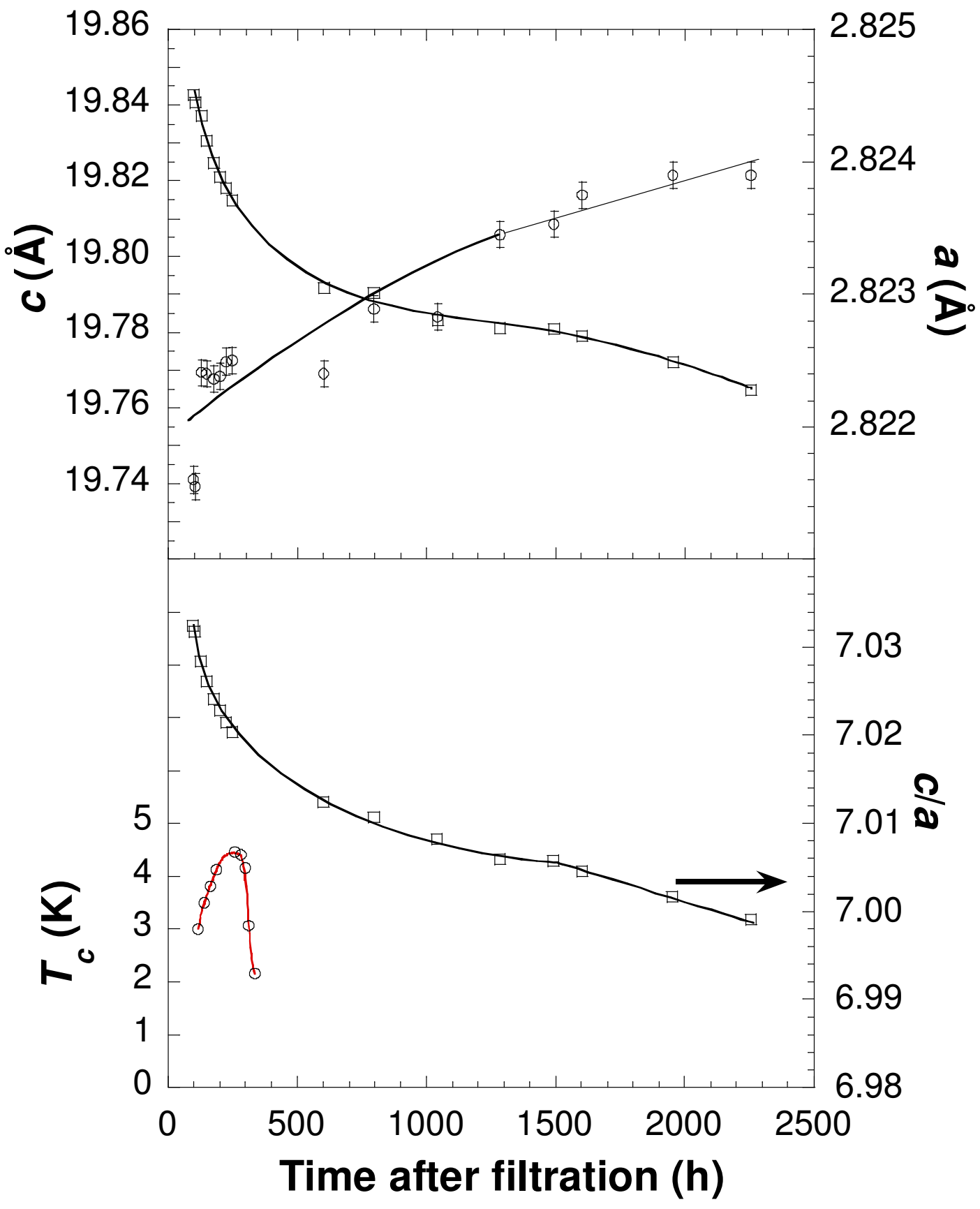

**Figure 7** Oxygen site occupancy and vacancies ($\delta$) in the $CoO_{2-\delta}$ layer vs. *c*/*a* for samples #4. The equation for the line shown is $\delta = 17.265 - 2.4965(c/a)$.

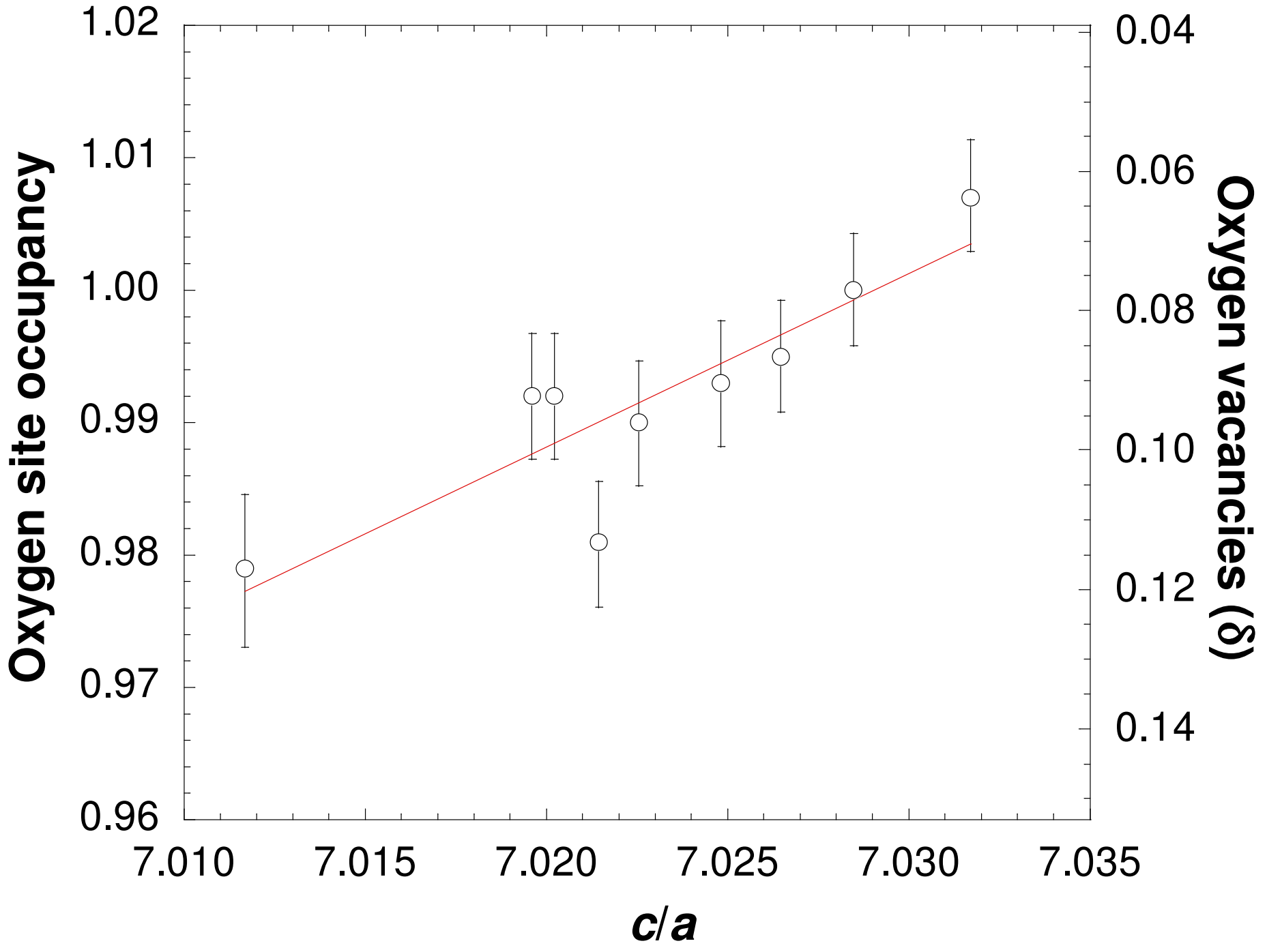

**Figure 8** Oxygen site occupancy vs. time after filtration for samples #4 (circles) and #5 (squares). The inset shows the changes in site occupancy over long time periods for sample #5. The lines shown here are drawn to guide the eyes. The shaded "superconductivity" region refers to the time scale for sample #4.

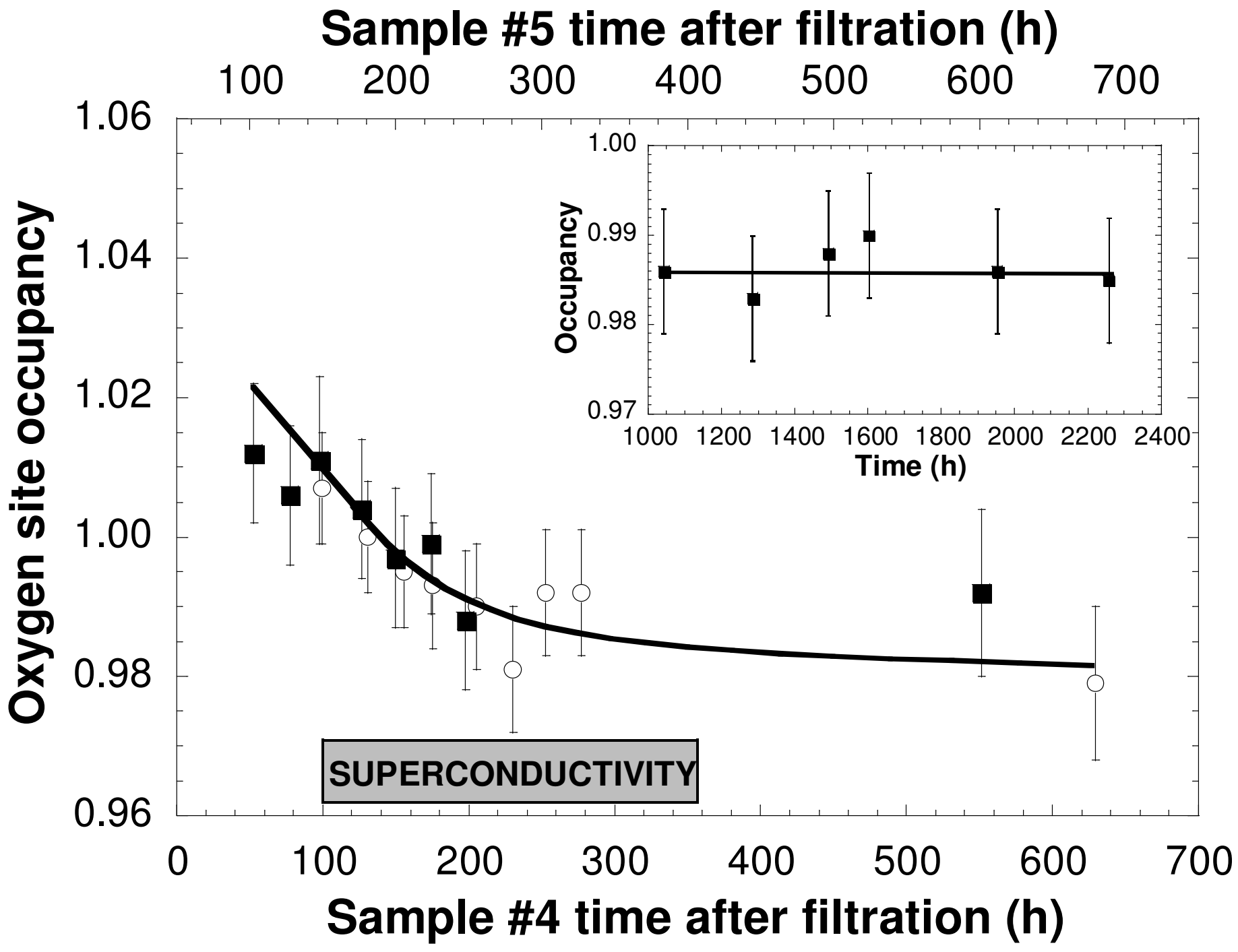

**Figure 9** Formal cobalt oxidization state plotted as a function of $T_c$ (onset) for sample #4. The line shown here is drawn to guide the eyes. The esd shown at $T_c$ = 4.45 K represents the error of the position of the dome with respect to cobalt oxidation state.

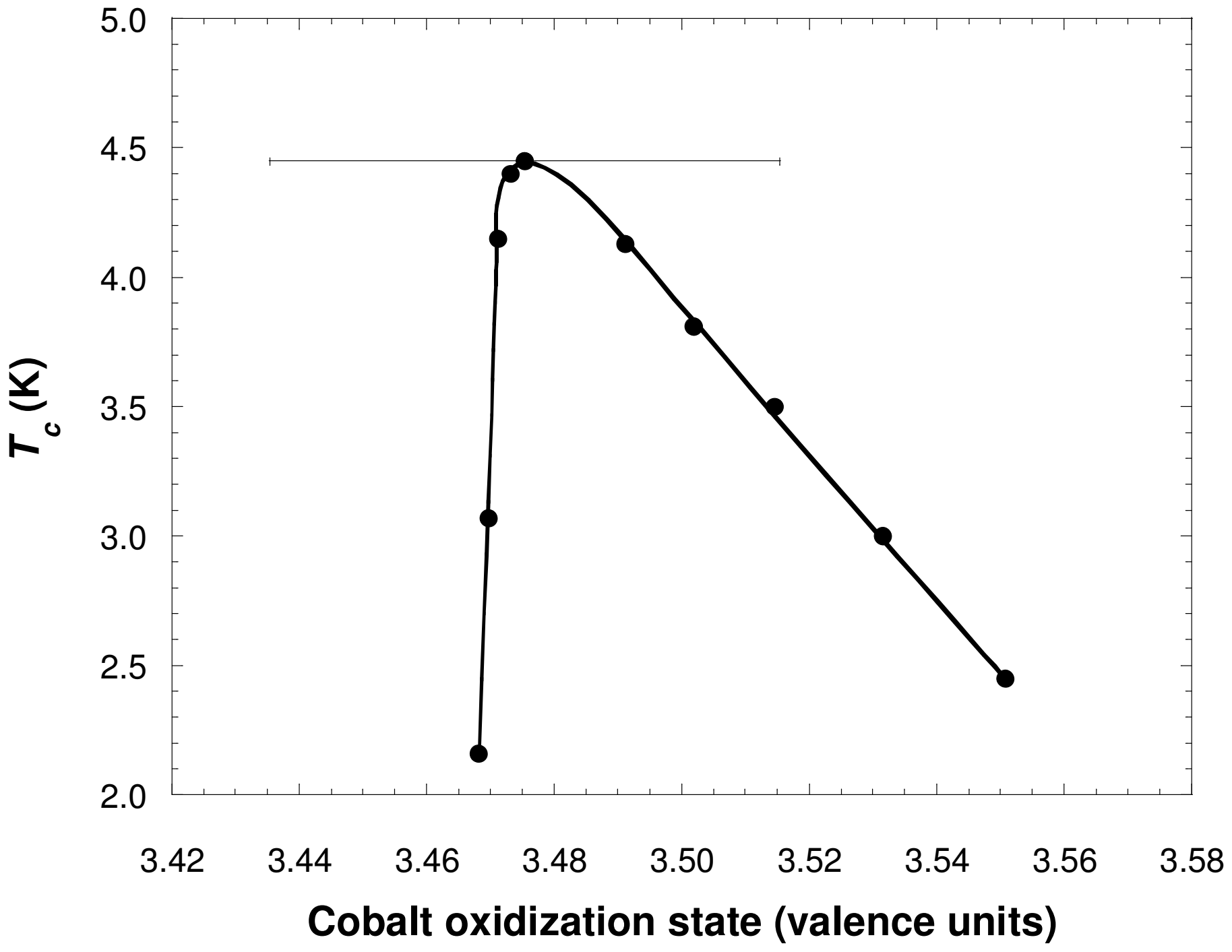